# On subwavelength imaging with Maxwell's fish eye lens

Fei Sun[1] and Sailing He[1,2]*

[1] Centre for Optical and Electromagnetic Research, JORCEP [KTH-ZJU Joint Research Center of Photonics], East Building #5, Zijingang campus, Zhejiang University (ZJU), Hangzhou 310058, China

[2] Department of Electromagnetic Engineering, School of Electrical Engineering, Royal Institute of Technology (KTH), S-100 44 Stockholm, Sweden

* Corresponding author: sailing@kth.se

**Abstract:** Both explicit analysis and FEM numerical simulation are used to analyze the field distribution of a line current in the so-called Maxwell's fish eye lens [bounded with a perfectly electrical conductor (PEC) boundary]. We show that such a 2D Maxwell's fish eye lens cannot give perfect imaging due to the fact that high order modes of the object field can hardly reach the image point in Maxwell's fish eye lens. If only zeroth order mode is excited, a subwavelength image of a sharp object may be achieved in some cases, however, its spot-size is larger than the spot size of the initial object field. The image resolution is determined by the field spot size of the image corresponding to the zeroth order component of the object field. Our explicit analysis consists very well with the FEM results for a fish eye lens. Time-domain simulation is also given to verify our conclusion. Multi-point imaging for a single object point is also demonstrated.

## 1. Introduction

Maxwell's fish eye was proposed by Maxwell in1854 [1]. Maxwell's fish eye gives a good image with equal light paths from the viewpoint of geometrical optics [1-3]. Recently, Leonhardt claimed that Maxwell's fish eye can give perfect imaging in wave optics and he modified the original fish eye lens, which is infinitely large, so that the device becomes finite [bounded with a perfectly electrical conductor (PEC) boundary] [4, 5]. Leonhardt gave an explicit solution with very small spot sizes of the object and image fields for such a fish eye lens with a line current source in the object point and a drain at the image point [4]. However, this configuration is not practical for imaging. For example, we do not know beforehand the distribution of fluorescent points in bio imaging, and thus we cannot determine where to put the drains in order to achieve an image of excellent resolution. If we put many drains beforehand, it may degrade the image resolution, particularly when some drains are located along the line connecting the object and the image. Apparently this is not a conventional concept for imaging. In a conventional image, we consider a very sharp field distribution (produced by some kind of source) and see if a lens can give a very sharp field distribution at another space point (without any drain). In this paper, we study the subwavelength imaging properties (in a conventional sense) of Maxwell's fish eye lens in the framework of wave optics. We show that perfect imaging can not be achieved due to the fact that high order modes of the object field will decay quickly before reaching the image point in Maxwell's fish eye lens. The image resolution is determined by the image field spot size corresponding to the zeroth order component of the object field and is related to the structure of Maxwell's fish eye and the location of the object. We find that for some cases the image resolution is diffraction-limited, while for some other cases it may still give a subwavelength image (without any drains introduced). We also study the influence of the radius of Maxwell's fish eye

(normalized to the wavelength) and the location of the object to the image resolution. Both explicit analysis and numerical simulation are given and they agree very well.

## 2. Mode analysis in Maxwell's fish eye lens

Maxwell's fish eye lens has the following refraction index profile [2]:

$$n = 2n_0 / [1 + (r/R_0)^2] \tag{1}$$

where $n_0$ and $R_0$ are the refraction index constant and radius of the reference sphere, respectively, and $r = \sqrt{x^2 + y^2}$ is the distance between a space point (x,y) and the center of Maxwell's fish eye lens. The Helmholtz equation for field $E_k$ corresponding to a source at point (0,0) (with a vacuum wave number $k=\omega/c=2\pi/\lambda$) in 2D space can be written as:

$$\frac{1}{r}\frac{\partial}{\partial r}(r\frac{\partial E_k(r,\theta)}{\partial r}) + \frac{1}{r^2}\frac{\partial^2 E_k(r,\theta)}{\partial \theta^2} + n^2 k^2 E_k(r,\theta) = g(r,\theta) \tag{2}$$

where $g(r,\theta)$ is the source term and $g(r,\theta)=0$ (when $r \neq 0$). Through variables separation $E_k(r,\theta)=U_k(r)\Theta(\theta)$, we can obtain the following general solution to Eq. (2)

$$E_k(r,\theta) = \sum_{m=0}^{+\infty}[a_m P_v^m(\zeta(r)) + b_m P_v^m(-\zeta(r))]e^{im\theta} \equiv \sum_{m=0}^{+\infty} E_k^m(r)e^{im\theta} \tag{3}$$

where

$$\zeta(r) = (r^2 - R_0^2)/(r^2 + R_0^2) \tag{4}$$

$$R_0^2 n_0^2 k^2 = v(1+v) \tag{5}$$

Here $P_v^m(\zeta(r))$ is the associated Legendre functions. We can see that the field distribution in Maxwell's fish eye lens can be expressed as a superposition of different order modes. m=0 and m$\neq$0 represent the zeroth order mode and the high order modes, respectively, and the high order modes correspond to high angular frequency components. Different sources may excite different modes.

## 3. Zeroth order mode in Maxwell's fish eye lens

In Maxwell's fish eye lens we first set a line current at point $(x_0,y_0)$ that can only excite the zeroth order mode. The Helmholtz equation for field $E_k$ in 2D space can then be written as:

$$\Delta_{x,y} E_k + n^2 k^2 E_k = \delta(x-x_0, y-y_0) \tag{6}$$

We shall consider Eq. (2) in domain D={(x,y)|$x^2+y^2<R_0^2$} and assume that $(x_0,y_0)\in$D. Let S={(x,y)| $x^2+y^2=R_0^2$} denote the boundary of D. Let function $E_k$ (x,y) satisfy on S the PEC boundary condition:

$$E_k|_S = 0 \tag{7}$$

Solution to the problem of Eqs. (6) and (7) gives Green's function to the Helmholtz equation with PEC boundary condition, and this can be easily found through the construction made by Leonhardt in [4], namely, through introducing complex plane z=x+iy and Möbius transformation:

$$w(z) = -z_\infty(z-z_0)/(z-z_\infty) \tag{8}$$

where $z_0 \equiv x_0+iy_0 = R_0\exp(i\chi)\tan\gamma$ and $z_\infty=-R_0^2/z_0^* = -R_0\exp(i\chi)\cot\gamma$. Furthermore, let $v=v(k)$ be a root of Eq. (5) and function $\xi(w(z))$ be determined by

$$\xi(w(z)) = (|w(z)|^2 - R_0^2)/(|w(z)|^2 + R_0^2) \qquad (9)$$

The solution to the problem of Eqs. (6) and (7) is given by

$$E_k(z) = [P_\nu(\xi(w(z))) - P_\nu(\xi(w(R_0^2/z^*)))]/4\sin\nu\pi \qquad (10)$$

where the intensity $P_\nu(\zeta)$ is the Legendre functions and $z^*=x-iy$. Indeed it was demonstrated in [4] that both $P_\nu(\xi(w(z)))$ and $P_\nu(\xi(w(R_0^2/z^*)))$ are solutions to the source-free Helmholtz equation for $z \neq z_0$ and, moreover, $P_\nu(\xi(w(z)))/4\sin(\nu\pi) \sim \ln|z-z_0|/2\pi$ as $z \to z_0$ and is a bounded smooth function outside a small vicinity of $z_0$, while the second term in (10) is a bounded smooth function everywhere in D [6]. Thus, $E_k(z)$ given by Eq. (10) fulfills the necessary singularity corresponding to a line current at $z_0$. On the other hand, on boundary S, we have $z=R_0^2/z^*$, and thus $E_k=0$.

Therefore, Eq. (10) gives a solution to the problem of Eqs. (6) and (7). Point $R_0^2/z_\infty^* \in D$ is the image of point $z_0$. All rays of Riemannian metric $ds=n(r)$ emitted from $z_0$ will be focused (after reflection on S) at the image point. This explains the fact that was noted numerically (see below) that the electric field has a local maximum at the image point. This may lead to subwavelength imaging of the source for some cases. For example, if we choose $R_0=5\lambda$ and $\lambda=0.2$m, and set a line current at $z_0(-0.5\text{m},0)$, we can use Eqs. (8), (9) and (10) to obtain the following field distribution in Maxwell's fish eye lens:

$$E_k(z) = [P_\nu(\frac{3r^2 + 8r\cos\theta - 3}{5r^2 + 5}) - P_\nu(\frac{-3r^2 + 8r\cos\theta + 3}{5r^2 + 5})]/4\sin\nu\pi \qquad (11)$$

The results of our analytical solutions (11) agree well with FEM simulation results as shown in Fig. 1. We can see the spot size [i.e., full width at half magnitude (FWHM)] around the image point is FWHM=$0.2925\lambda$ (indicating subwavelength image), which is larger than the spot size FWHM=$0.1825\lambda$ around the source point. For comparison, we also show in Fig. 1 Leonhardt's analytical solution for a special situation when one sets a drain at the image position with the same intensity of the original source of line current [4]:

$$E_k(z) = \frac{1}{4\sin(\nu\pi)}\{[P_\nu(\xi(z)) - P_\nu(\xi(R_0^2/z^*))] - e^{i\pi\nu}[P_\nu(-\xi(z)) - P_\nu(-\xi(R_0^2/z^*))]\} \qquad (12)$$

The solution (12) given in [4] corresponds to a linear combination of delta-function sources localized at 2 points: $z_0$ (source location) and $R_0^2/z_\infty^*$ (drain location) inside the PEC boundary (equivalent to a linear combination of delta-function sources localized at 4 points: $z_0$, $z_\infty$, $R_0^2/z_0^*$ and $R_0^2/z_\infty^*$ in the whole unbounded space), and is not a Green's function for Eqs. (6) and (7). From Fig. 1 we see that Maxwell's fish eye lens can still give a good image of subwavelength resolution if only the zeroth order mode is excited without any drain. However, the spot-size of the image field is still larger than the spot size of the initial source field (indicating that it can not give a perfect image). Adding a drain at the image point [4] may sharpen the image spot size, and even recover the object shape for a very special excitation of object field of only zeroth order mode. However, it is not practical to add drains in a real imaging application, as mentioned earlier. Furthermore, a simple line drain can not produce enough high order modes to make the image as sharp as one wishes (for perfect image) though it can help to recover roughly a very special object field distribution (of a finite spot size and zeroth order mode) around the image position. We will not discuss the situation of drains in this paper.

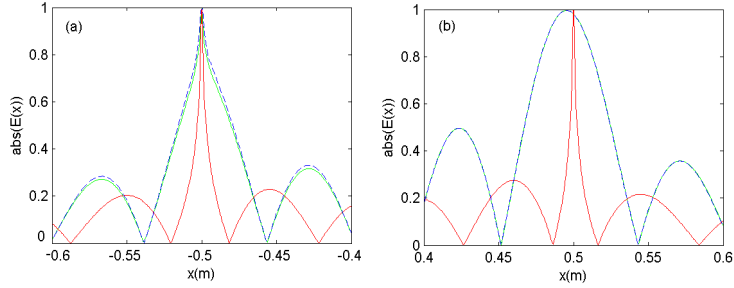

Fig. 1. The absolute value of the normalized field distribution around the line current (a) and its image (b) along x direction: blue dashed line is from the FEM simulation result when we set a line current at (-0.5m, 0); green line is our analytical result of Eq. (11); red line is Leonhardt's analytical result of Eq. (12) for a situation when one sets a line current source at (-0.5m,0) and a drain at (0.5m,0). The parameters for the fish eye lens are $R_0=5\lambda$ and $n_0=1$. The incident wavelength is $\lambda=0.2m$.

We should note that for a given structure of Maxwell's fish eye lens, if the position of the line current changes, the spot size around the image point will also change. The results are shown in Fig. 2. As $abs(x_0)/\lambda$ increases, the spot size (FWHM) of the image has an over-all increasing trend (as the refractive index at the image point becomes larger), however, with some small oscillating behavior locally (due to the introduction of the PEC boundary, as explained at the end of Appendix). If we increase the size of the fish eye lens without changing the normalized position of the line current, the spot size of the image field around the image point will decrease. Thus we can see for some cases the image resolution in Maxwell's fish eye lens is diffraction-limited, while for some other cases it may still be able to give a subwavelength image (without any drains introduced).

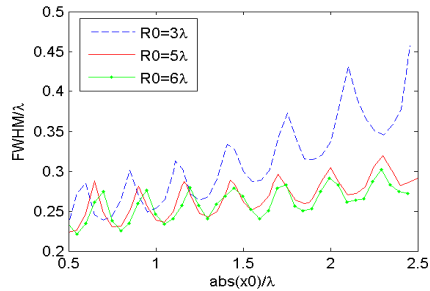

Fig. 2. Spot size of the image field around the image point when the position of the line current varies. The horizontal axis indicates the normalized position of the image. Lines of different colors indicate different sizes of the fish eye lens.

We know that one can obtain two kinds of Green functions by solving the stationary wave equation in Maxwell's fish eye medium filled in the whole space without any boundary. One is the retarded Green function which is casual, and the other is the advanced Green function which is not causal [7]. Only the retarded Green function is physically meaningful in Maxwell's fish eye medium filled in the whole space without any boundary. However, when we set a PEC boundary in Maxwell's fish eye medium, the advanced Green function is associated to the wave reflected from the PEC boundary and thus is also meaningful. The field distribution produced by a line current in Maxwell's fish eye with PEC boundary should be the superposition of an advanced Green function and a retarded Green function. Our analytical solution Eq. (10) is therefore causal and meaningful. To verify that our analytical solution is causal, we made the following FDTD simulation: We set a line source with a single frequency ($\lambda=0.2m$) at position (-0.5m,0) to produce a continuous wave (excitation field, but not the total field) in a 2D fish eye bounded with PEC at a radius of 1m. The simulation result is shown in Fig. 3. The electric field propagates from the source to the image point and starts to form a good image at time t=10.3333ns. Then the field forming the sharp image will behave like a new source and propagate back to the source point forming a new source due to the "confocus" property of the lens bounded with PEC. This process repeats again and again. After about 42ns, we find the field in Maxwell's fish eye keeps the harmonic oscillation for quite a long time. The normalized integration of the absolute value of the electric field over one time-harmonic period (indicating the local magnitude of the field) is shown in Fig. 3(a). From this field distribution one sees that the spotsize around the image is FWHM=0.2900λ (consistent with our earlier analysis in frequency domain), which is bigger than the spotsize around the object FWHM=0.2538λ. We note that the spotsize around the object is bigger than our earlier analysis in frequency domain. The reason for this is that the source in our FDTD simulation is a soft source which keeps on producing the incident field of fixed magnitude (but can not keep the total field fixed at the source point, unlike our earlier FEM simulation). Due to the soft source in our simulation and the PEC boundary of Maxwell's fish eye lens with lossless medium, the total energy in the whole Maxwell's fish eye lens increases with time. According to our calculation, the total average magnitude in one harmonic-time period at t=333.3333ns is about three time larger than that at t=41.6667ns. This will also lead to large "crosstalk" (adjacent peaks) of the object and image as time increases further (see Fig. 3(b)). This may be due to the above-mentioned repeated confocusing phenomenon.

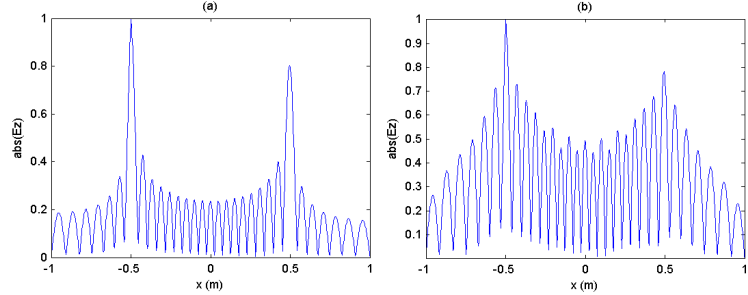

Fig. 3. The normalized integration of the absolute value of the electric field over one time-harmonic period. The results are calculated with the FDTD method in 2D Maxwell's fish eye bounded with the PEC at the radius of 1m, and plotted along the straight line passing both the source and image points. We set a line source at position (-0.5m, 0) to produce a single frequency wave with λ=0.2m. (a) During one period t=41.6667ns~42.3333ns. The spotsize around the image is FWHM=0.2900λ, which is larger than the spotsize around the object FWHM=0.2538λ. (b) During another period t=333.3333ns~334.0000ns. The object and image have a large crosstalk (adjacent peaks).

To shed more light on the imaging performance of Maxwell's fish eye bounded with PEC, we make some additional numerical simulations in time domain with the FDTD method. We set a line source at position (-0.5m,0) to produce a narrowly localized Gaussian wavepacket with pulse function $J(t)=\exp[-(t-t_0)^2/\Delta^2]\cos[\omega(t-t_0)]$ in the 2D fish eye bounded with the PEC at the radius of 1m. We choose pulse width $\Delta$=0.0167ns, $t_0$=0.3333ns and $\lambda_c=2\pi c/\omega$=0.2m. The simulation results are shown in Fig. 4, from which we can see that a wavepacket can be formed around the image point (0.5m,0). When the pulse field reaches its maximum at the image point, the spot size is FWHM=0.1505$\lambda_c$ (see Fig. 4(e)), which is much larger than the initial spot size FWHM=0.0310$\lambda_c$ around the source point (see Fig. 4(b)). Thus, a narrow wavepacket cannot give an equally narrow focus at the image point in a 2D Maxwell's fish eye lens.: Then the electrical field around the image location starts to decreases as it propagates back to the source location forming a new peak there with FWHM=0.0830$\lambda_c$ (at time t=22.0580ns, see Fig. 4(f)) due to the confocus property of the lens bounded with PEC.

According to our earlier analysis, even for a time-harmonic line current which can only produce single-frequency zeroth order mode field in the 2D fish eye bounded with PEC, we cannot obtain an equally sharp field spot at the image point (see Fig. 1) as compared to the initial source field. Since a pulse wavepacket of source field contains many frequency components, different frequency components form image spots of different sizes (as we have explained earlier, see Fig. 2). Consequently, the superposition of different frequency components at the image point will form a wavepacket of larger spot size as compared to the spot size of the initial source field (a narrowly localized wavepacket).

This time-domain simulation result, which is obviously causal, is consistent with our earlier frequency-domain analysis in the present paper: When one sets only a line current without any drain in 2D Maxwell's fish eye with PEC boundary, one can still obtain a subwavelength image spot which, however, is wider than the initial sharp spot size around the source point.

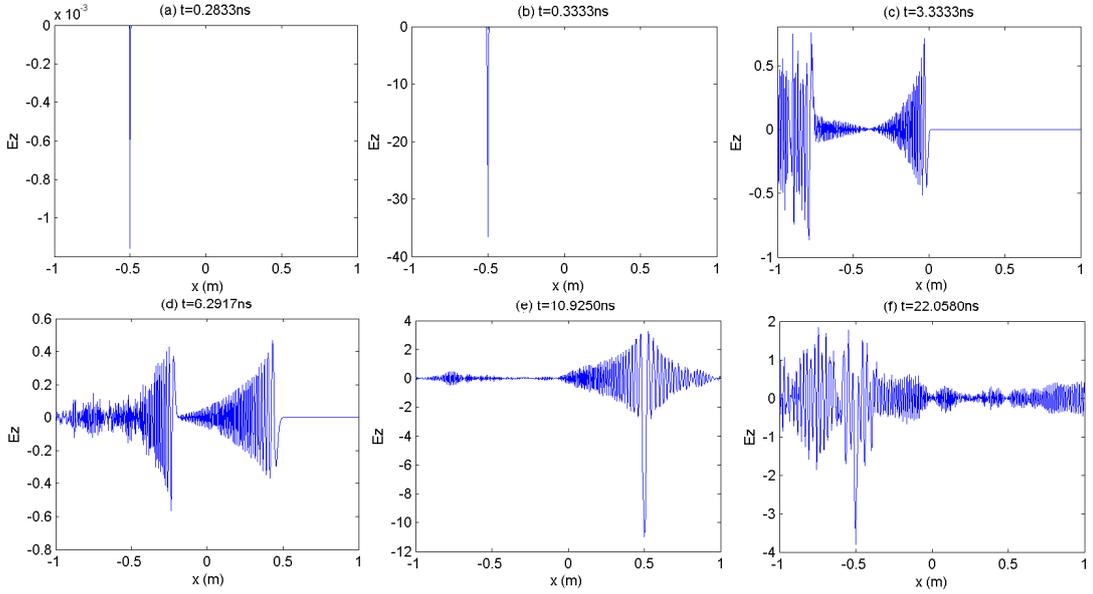

Fig. 4. Electrical field distribution (along the straight line passing both the source and image points) calculated with the FDTD method at different times in 2D Maxwell's fish eye bounded with the PEC at the radius of 1m. We set a line source at position (-0.5m,0) to produce a Gaussian wavepacket with time-varying function $J(t)=\exp[-(t-t_0)^2/\Delta^2]\cos[\omega(t-t_0)]$, where $t_0=0.3333$ns, $\Delta=0.0167$ns and $\lambda_c=2\pi c/\omega=0.2$m. (a) At time t=0.2833ns the electric field starts to appear around the source location. (b) At t=0.3333ns the pulse field reaches its maximum around the source location with spatial spot size FWHM=0.0310$\lambda_c$. (c) At t=3.3333ns: the electric field propagates from the source toward the image. (d) At t=6.2917ns, the electric field just reaches the image location. (e) t=10.9250ns: a good image is formed and the electrical field at the image location reaches its maximum with a spatial spot size FWHM=0.1505$\lambda_c$. (f) t=22.0580ns: the electrical field around the image location decreases as it propagates back to the source location forming a new peak there with FWHM=0.0830$\lambda_c$ due to the confocus property.

### 4. Case for object fields with high order modes

In this section, we study numerically (instead of analytically) the propagation of high order modes in Maxwell's fish eye. First we show that if we put at the center of the original Maxwell's fish eye (without PEC) a source (e.g. $\delta(r)f(\theta)$) that can produce high order mode of angular momentum, we cannot get an image spot for those high order modes. The dispersion relationship in a cylindrical coordinate system whose origin is located at the center of Maxwell's fish eye can be written as [8]:

$$k_r^2 + k_\theta^2 = n^2 k^2 \tag{13}$$

where $k_r$ is the radial component of the wave vector, and $k_\theta$ is the tangential component of the wave vector. Considering the conservation of angular momentum for the m-th order mode, we have

$$rk_\theta = m \tag{14}$$

When m≠0, from Eq. (14) we can see that $k_\theta$ increases toward the center. Consequently, we can see from Eq. (13) that radial component $k_r$ varies from a real value to an imaginary value as r→0. The turning point of $k_r$=0 is the radius of the caustic. Inside the caustic, $k_r$ is an imaginary number and the angular momentum state becomes evanescent (i.e., decays quickly) along the radial direction. The detailed information carried by the high order modes can hardly propagate to the far field without great damping. Only the zeroth order mode (m=0), which does not have the caustic, can propagate to the far field in Maxwell's fish eye. Thus, if we put at the center of Maxwell's fish eye a special source that can excite only (or mainly) high order mode, the field cannot go to the far field, and consequently a subwavelength image can not be formed. If we transform this source position to another point of Maxwell's fish eye or add PEC boundary to Maxwell's fish eye, the situation remains the same: subwavelength image can not be achieved.

We can use FEM simulation to verify this in Maxwell's fish eye with PEC boundary at $R_0$=10λ and $n_0$=1. Our simulation is for TE wave in 2D space with λ=0.2m. We set a small circle (with radius $r_0$) located at $z_0$(-0.5,0) with boundary condition E=3exp(iγθ') V/m to introduce some high order mode. We first choose $r_0$=$10^{-3}$λ and γ=0 and the simulation result is shown in Fig. 5. Note that the field generated by boundary condition E=3 V/m on this small circle will contain some high order mode (and thus the object field is quite sharp as compared to Fig. 1(a)), as the zeroth order mode produced by a line current at (-0.5m, 0) is not a circle (see Appendix). Since it also contains some zeroth order mode, a good image spot can still be formed. However, if we change γ=0 to γ=5, the situation will be completely different. The simulation result is shown in Fig. 6. Boundary condition E=3exp(i5θ') V/m on a small circle gives more energy to high order modes (the object field is very sharp in Fig. 6(a)). These high order modes cannot propagate to the far field (the ratio of the field magnitude around the object to that around the image position is about $E_o/E_i$~105). Consequently, good image can not be achieved, as shown in Fig. 6(b).

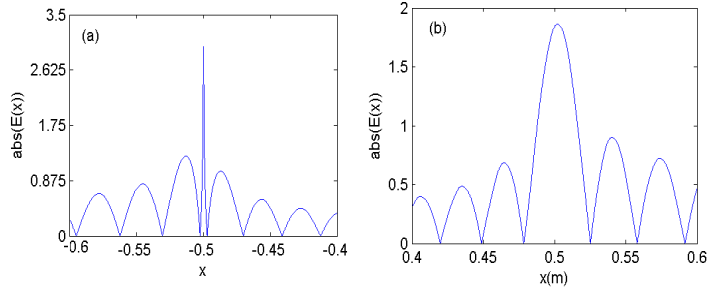

Fig. 5. FEM simulation results for the absolute value of the field distribution around the object (a) and its image (b) along x direction for Maxwell's fish eye lens with $R_0$=5λ and $n_0$=1. The object field is excited with boundary condition E=3exp(iγθ') V/m at a small circle located at (-0.5m,0) with γ=0 and $r_0$=$10^{-3}$λ. Here we choose λ=0.2m.

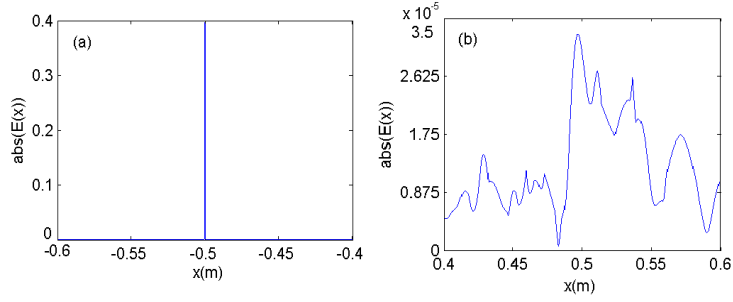

Fig. 6. FEM simulation results for the absolute value of the field distribution around the object (a) and its image position (b) along x direction for the same Maxwell's fish eye lens. We have set γ=5 (while keeping the other parameters the same as those for Fig. 5 to excite more energy to high order modes.

## 5. Multi-point imaging in Maxwell's fish eye lens

We find that if the radius of PEC boundary R does not equate to the radius of the reference sphere $R_0$, some interesting phenomenon may happen. Fig. 7 shows that multi-point imaging can be formed when we set a line current in a special structure of Maxwell's fish eye lens with radius of PEC boundary R=10λ and the radius of the reference sphere $R_0$=5λ. This phenomenon may have some other applications such as multi-point laser direct writing in parallel.

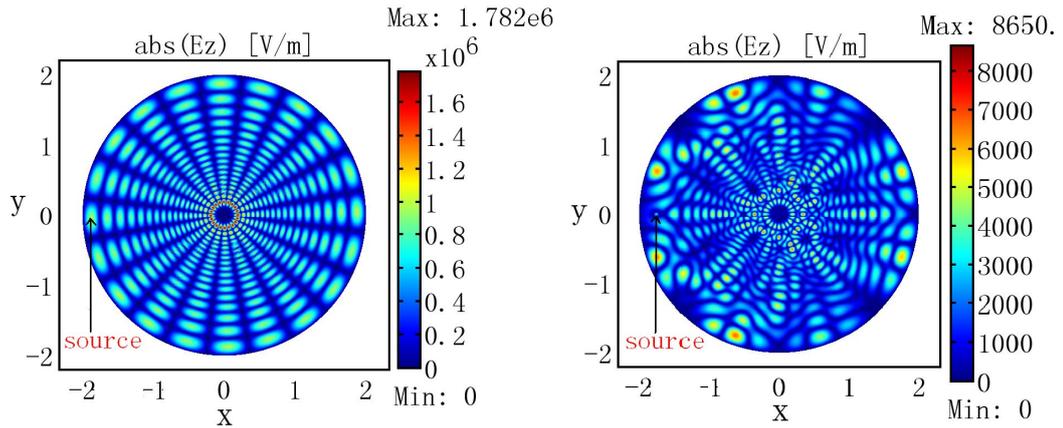

Fig. 7. FEM simulation results for the absolute value of the field distribution in the modified fish eye with R=10λ, $R_0$=5λ, λ=0.2m, and $n_0$=1 (a) when we set a line current at $z_0$ (-1.85m,0); (b) when we set a line currents at $z_0$ (-1.75m,0)

## 6. Conclusions

Maxwell's fish eye lens of some special structures can give a good image of subwavelength resolution (less than 0.3 wavelength) for a line current (without any drain) that excites only zeroth order mode. However, as we have shown in the present paper, such a Maxwell's fish eye lens cannot give perfect imaging since high order modes of the object field are evanescent modes and can hardly reach the far-field image point. Good image can not be achieved when the object field contains mainly high order modes. The image resolution is determined by the field spot size of the image corresponding to the zeroth order component of the object field. Both explicit analysis and FEM numerical simulation have been performed and they agree very well with each other. The dependence of the spot size of the image on the position of the line current and the lens size has also been given. Time-domain simulation has also been carried out and the numerical results are consistent with our analysis.

**Acknowledgments**

An earlier version of this work can be found on ArXix: http://arxiv.org/abs/1005.4119, which was submitted on May 22, 2010. The authors are grateful to Prof. Vladimir Romanov for many valuable discussions and Xiaocheng Ge for great help in numerical simulation. We also thank Dr. Yi Jin, Pu Zhang, Yuqian Ye, Yingran He, and Jianwei Tang for some helps. The work is partly supported by the National Basic Research Program, the National Natural Science Foundations of China, and the Swedish Research Council (VR) and AOARD.

**Appendix**

From the viewpoint of transformation optics [4], we know if we set a line current at the North Pole N on a spherical surface an image spot can be formed at the South Pole S on the spherical surface. According to the symmetry, the field produced by a line current and its image should be the zeroth order mode of circular symmetry on the spherical surface. When we make a transformation from a spherical surface to a plane, the electric field distribution will also be transformed. The zeroth order mode on the spherical surface is also transformed into the zeroth order mode in Maxwell's fish eye. We can use a stereographic projection [4] to transform a spherical surface to a plane. However, when the zeroth order mode is centered at different positions on the spherical surface, we have different projection shapes on the plane. That is the reason why we have different shapes of the zeroth order mode at different places of Maxwell's fish eye and the modified one bounded with the PEC. A circle on a spherical surface may be transformed to an ellipse on the plane. We assume the radius of the zeroth order mode around the line current or its image on the spherical surface is $R_{zero}$. Considering the symmetry of a spherical surface, we can assume the center of this zeroth order mode is on the x-z plane. If we make a stereographic projection of this circle on the spherical surface to the plane, we will obtain an ellipse function:

$$\frac{(x-x_0)^2}{a^2} + \frac{y^2}{b^2} = 1 \tag{A1}$$

where

$$a = \sqrt{(D + \frac{C^2}{4B})/B}, \ b = \sqrt{(D + \frac{C^2}{4B})/A}, \ x_0 = -\frac{C}{2B} \tag{A2}$$

and where

$A = (\cos\theta_0 + \sqrt{R_0^2 - R_{zero}^2})^2, \ B = (1 + \cos\theta_0\sqrt{R_0^2 - R_{zero}^2})^2 - R_{zero}^2 \sin^2\theta_0,$

$C = 2R_0^2 \sin\theta_0 \cos\theta_0 + 2\sin\theta_0\sqrt{R_0^2 - R_{zero}^2}, \ D = R_{zero}^2 - R_0^2 \sin^2\theta_0.$

Here $R_0$ is the radius of the reference sphere, $R_{zero}$ is the radius of the zeroth order mode around the line current or its image on the reference sphere, $(x_0,0)$ is the center of the projected elliptic disk on the plane around the line current or its image. $\theta_0$ is the angle between the z axis and the line connecting the center of the zeroth order mode and the origin (see Fig. A1).

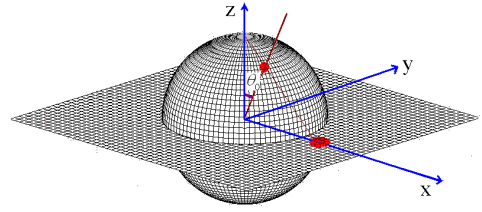

Fig. A1. Stereographic projection. The zero order mode with circular symmetry on the spherical surface will be projected into a modal field of ellipse shape on the 2D plane.

If we know the radius of the zeroth order mode on the spherical surface (denoted by $R_{zero}$), we can use Eq. (A2) to calculate the size of the zeroth order mode in 2D Maxwell's fish eye. Let a and b denote the half widths along the x and y directions, respectively. According to Eqs. (A1) and (A2), we can see if the object is near the original point (i.e., $\theta_0$ is small) and $R_{zero} \ll R_0$, the zeroth order mode is of circular shape (a~b). If the object is far from the origin (i.e., $\theta_0$ is large), the zeroth order mode will become an ellipse (a≠b). Note that the projected spot on the plane will be inside the circle with radius $R_0$ when the original field spot is on the lower surface of the sphere. For Maxwell's fish eye lens of a specific size, $R_0$ and $R_{zero}$ are fixed. From (A2) we can see when we change $\theta_0$, both a and b will change, and may reach maximum at some special $\theta_{0m}$. For Maxwell's fish eye lenses of different sizes, $\theta_{0m}$ also differs as one can see from Eq. (A2).

If we add a PEC boundary at the equator of the sphere, the symmetry (about angle $\theta_0$) of the spherical surface has been broken. Thus, image field spot size $R_{zero}$ (produced by a line source) should also depend on the position angle $\theta_0$. Consequently, the projected elliptic disk on the plane will also have different size. This explains the small oscillating behavior in Fig. 2.